# Structural and electromechanical characterization of lead magnesium niobate-lead titanate (PMN-0.3PT) piezoceramic for energy harvesting applications

Abhishek Kumar, Amritendu Roy*

School of Minerals, Metallurgical and Materials Engineering, Indian Institute of Technology Bhubaneswar, Odisha-752050, India

*Corresponding author: amritendu@iitbbs.ac.in

## Abstract

Efficient mechanical energy harvesting using the principle of piezoelectric effect demands specific material-property requirements. This includes a combination of large piezoelectric charge coefficient ($d_{ij}$), large elastic strain ($\varepsilon_y$), small elastic compliance ($S_{ij}$), and small dielectric permittivity ($\kappa_{ij}$). The present work undertakes structural, electrical, mechanical, and electromechanical characterization of pyrochlore-free lead magnesium niobate-lead titanate $(1-x)[Pb(Mg_{1/3}Nb_{2/3}O_3)] - xPbTiO_3$ at $x = 0.3$ or PMN-0.3PT, to estimate the above critical parameters for mechanical energy harvesting. Pyrochlore-free PMN-0.3PT ceramic with co-existing monoclinic (*Pm* and *Cm*) phases was synthesized using solid-state reaction method. Piezoelectric charge coefficient ($d_{33}$), dielectric permittivity ($\kappa_{33}^T$), elastic compliance ($s_{33}^E$), and electromechanical coupling factor ($k_{33}$), were estimated to be, 200 ± 21 pC/N, 1.06 (± 0.06) × 10$^{-8}$ F/m, 13.16 (± 0.2) ×10$^{-12}$ m$^2$/N, and 0.54 ± 0.04, respectively, using room temperature impedance measurement on a poled sample with specified dimensions (EN 50324-1:2002 and CEI/IEC 60483:1976). Polarization leakage due to transport of various charged defects was identified to be responsible for the reduced electromechanical properties compared to those reported for single crystals. Elastic strain ($\varepsilon_y$) vis-à-vis flexibility ($f_{FOM}$) of the PMN-0.3PT was estimated to be 4.5 ×10$^{-4}$. Energy harvesting under dynamic mechanical loading shows a maximum short-circuit current density, 95 nA/cm$^2$, and an open-circuit electric field, 98 V/cm. With its impressive performance, PMN-0.3PT ceramic constitutes an important material for piezoelectric energy harvesting.

## 1. Introduction

Piezoelectric generators (PGs), also termed piezoelectric energy harvesters (PEHs) are solid-state devices designed to convert mechanical excitation into electrical power using the principle of direct piezoelectric effect [1]. Over the last two decades, PGs have attracted considerable attention because of their ability to harvest ubiquitous mechanical energy from ambiance and power other devices with small energy requirements, especially those inaccessible to the grid power and/or not ideal for battery operation. PGs could be, thus, employed for various types of devices, such as implantable wireless sensors, MEMS systems, and wearable electronic devices, without the challenge of replacing and charging batteries [2,3].

PGs typically convert the available ambient mechanical energy such as vibration (vehicle engine, pumps/motors, etc.), bending or tension (wearable devices), and compression (tire-road interaction, walking or running on the pavement, etc.) into electricity. Based on the various mechanical energy sources, several device designs have evolved, namely cantilever-type [4], circular diaphragm [5], cymbal type [6], and stack type [7]. All these configurations work on different modes of input excitation. For instance, cantilever-type PGs work in the flexural mode ($d_{31}$ mode) [4], cymbal PGs operate on the radial mode (combined $d_{31}$ and $d_{33}$ modes) [6], circular diaphragm operates in $d_{31}$ mode [5] and stack PGs work on the compression mode ($d_{33}$ mode) [8]. Out of these configurations, stack-type PGs are expected to withstand large mechanical loads [9] and when suitably designed, can yield large power density.

Materials for PGs are designed based on the figure-of-merit analyses [10]. Out of several figure-of-merits, FOM 1 [10] and FOM 2 [11] are most relevant for devices under compression mode of operation: $FOM\ 1 = k_{ij}^2 = d_{ij}^2/\kappa_{ii}\ S_{jj}$ and $FOM\ 2 = d_{ij}^2/\kappa_{ii}$ where $k_{ij}$, $d_{ij}$, $\kappa_{ii}$, and $S_{jj}$ are the electromechanical coupling factor, piezoelectric charge coefficient, dielectric permittivity, and elastic compliance of the material, respectively. Both figure-of-merits suggest that piezoelectrics must have a large piezoelectric charge coefficient for high conversion efficiency, whereas dielectric permittivity and elastic compliance should be minimum. Among the various piezoelectric materials reviewed, relaxor ferroelectric $(1-x)Pb(Mg_{1/3}Nb_{2/3}O_3) - xPbTiO_3$ (PMN-PT) is a suitable piezoelectric due to its giant piezoelectric coefficient leading to large FOM 1 and FOM 2 (Supplementary section. Figure S1(a-b)).

Earlier reports on the energy harvesting performance of PMN-PT (single-crystal) cantilever-shaped vibrational energy harvester reported a maximum power of 0.3 mW under vibration mode of input excitation ($f = 1.3$ kHz) [12]. Energy harvesting with (001)-orientated PMN-0.28PT single-crystal reported an open-circuit voltage and short-circuit current of 8.2 V and 223 µA (working area of 2.89 cm$^2$) [13]. Tang *et al.* [14] developed a cantilever type piezoelectric energy harvester made of single-crystal PMN-PT (film) and the silicon layer. Their experimental results reported an open-circuit voltage of 5.36 V and output power of 7.182 µW at $f_r = 406.0$ Hz (resonance frequency). PMN-0.33PT single-crystal reported a maximum power density of 16.7 nW/(g$^2$ cm$^3$) under a resonance vibration ($f_r$) of 50 Hz [15]. PMN-PT nanobelt (single crystal) arrays fabricated by the top-down technique reported maximum current and output voltage of 102 µA and 6 V, respectively, under stretching and compression modes [16]. However, the synthesis of such piezoelectric single crystalline materials is quite expensive and complex; thus, piezoelectric bulk ceramics would be preferred from application perspectives. Rakbamrung et al. [17] reported an output power, 4.5 µW, in PMN-0.25PT ($t = 1$ mm, $d = 14$ mm), under vibration mode at high frequency. However, such high input frequency is rare in ambient. Therefore, the performance of PMN-PT-based PGs at low (off-resonance) frequency is vital. Furthermore, electromechanical properties of piezoelectric solid solutions are pronounced in the morphotropic phase boundary (MPB) regions (PMN-0.27-0.34 PT) [18]. Thus, it would be interesting to evaluate the energy harvesting performance of PMN-PT ceramic within the MPB region at a lower excitation frequency in compression mode.

In the present work, we selected a composition within the morphotropic phase boundary region and synthesized PMN-0.3PT piezoceramic using a low-cost, solid-state reaction route. Using a variety of structural, electrical, and electromechanical characterization techniques, PMN-0.3PT piezoelectric ceramic was studied. The piezoelectric charge coefficient ($d_{33}$), dielectric permittivity ($\kappa_{33}^T$), elastic compliance ($s_{33}^E$), and electromechanical coupling factor ($k_{33}$), were estimated to be, 200 ± 21 pC/N, 1.06 (± 0.06) × $10^{-8}$ F/m, 13.16 (± 0.2) ×$10^{-12}$ m$^2$/N, and 0.54 ± 0.04, respectively. Conduction due to the motion of doubly ionized oxygen vacancies, single ionized oxygen vacancies, and the hopping of charge carriers at different temperature ranges is responsible for the reduced polarization vis-à-vis electromechanical properties of PMN-0.3PT ceramic with respect to those of single-crystals. Energy harvesting performance of the PMN-0.3PT piezoceramic under cyclic compressive loading yielded an open circuit electric field and average output power density, 98 V/cm and ~ 9 µW/cm$^3$, respectively.

## 2. Experimental details
### 2.1 Synthesis of PMN-0.3PT ceramic

PMN-0.3PT ceramic was synthesized using solid-state reaction method. To start with, measured amount of commercially available MgO (5 wt.% excess, Sigma-Aldrich, ≥99%) and Nb$_2$O$_5$ (Sigma-Aldrich, 99.99%) powders were mixed in isopropyl alcohol using a roller pot-mill for 24 h. Yttria-stabilized zirconia (YSZ) balls of different sizes (1 mm-10 mm) were used with a ball to powder ratio of 6:1. The wet powder was dried, calcined at 950°C for 4 h to form magnesium niobate ($MgNb_2O_6$):

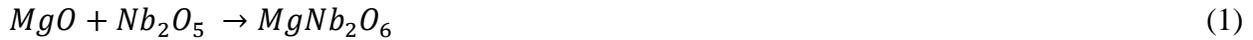

$$MgO + Nb_2O_5 \rightarrow MgNb_2O_6 \qquad (1)$$

Calcined powder was crushed and ball-milled with YSZ balls in isopropyl alcohol for another 48 h. Thereafter, PbO (25 wt.% excess, Sigma-Aldrich, ≥99%) and TiO$_2$ (Himedia, 99%) were added to the calcined $MgNb_2O_6$ powder and milled (in isopropyl alcohol) for another 48 h. The dried powder mix was subsequently subjected to another round of calcination at 950ºC for 4 h:

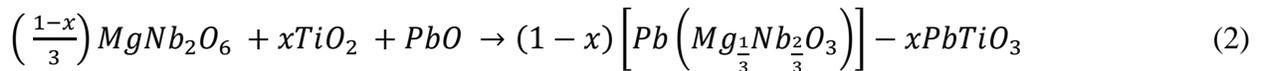

$$\left(\frac{1-x}{3}\right) MgNb_2O_6 + xTiO_2 + PbO \rightarrow (1-x)\left[Pb\left(Mg_{\frac{1}{3}}Nb_{\frac{2}{3}}O_3\right)\right] - xPbTiO_3 \qquad (2)$$

Calcined powder was crushed and milled again for 24 h. The resultant powder was pelletized and sintered at a constant heating rate of 5°C/min at 1250ºC for 4 h to form a pyrochlore-free phase of PMN-0.3PT.

### 2.2 Characterization

Structural characterization using powder x-ray diffraction was performed in a Bruker D8 Advance x-ray diffractometer with CuKα radiation (λ = 1.5418 Å). Microstructural characterization of the thermal-etched pellet was carried out with a field emission scanning electron microscope (Zeiss

Merlin compact) equipped with Energy Dispersive Spectroscopy (EDS). The hardness and fracture toughness of the well-polished PMN-0.3PT pellets were evaluated by Vickers micro-indentation testing (Zwick-Roell Indentec). The indentation load of 98 mN and 980 mN were used with a holding time of 10 s for determining the hardness and fracture toughness, respectively. Silver conductive paste was applied on two flat surfaces of the sample for electrical measurement. Impedance studies were carried out using Wayne Kerr 6500B impedance analyzer over a frequency and temperature range, 20 Hz to 1 MHz and 30 °C-425 °C, respectively. The sample was subsequently poled at 10 kV/cm for 30 min. at 80°C. Electromechanical and energy harvesting performance of the sample was evaluated upon poling. Energy harvesting performance was measured on an indigenously designed facility by applying a cyclic compressive load with an Arduino-operated stepper motor. The output current and voltage were measured using a Keithley 6517B high-resistance electrometer.

## 3. Results and discussion
### 3.1 Structural and microstructural studies

Figure 1 (a) plots room temperature powder x-ray diffraction (p-XRD) pattern of the as-synthesized PMN-0.3PT sample. The pattern was compared and indexed using ICSD database card no. 01-075-7998 and 01-075-8005 corresponding to *Cm* and *Pm* phases, respectively. No trace of additional phases such as pyrochlore was detected. Rietveld refinement was performed using three different structural models: Monoclinic *Cm* (space group no. 8), monoclinic *Pm* (space group no. 6), and co-existing *Cm* and *Pm* [18]. Figure 1 (a) reveals the profiles obtained by Rietveld refinement of the p-XRD data by considering *Cm*, *Pm*, and (*Cm* + *Pm*) phases, respectively. Typically, *R*-factors (statistically expected, $R_{exp}$, weighted profile, $R_{wp}$, and goodness-of-fit, $\chi^2 = (R_{wp}/R_{exp})^2$ are the key parameters for evaluating the quality of refinement [19]. Refinement considering *Cm* and *Pm* space groups, separately, shown in the top two panels of figure 1 (a), demonstrates a noticeable mismatch among the observed and calculated peak profiles, especially for the 200 and 210 reflections. This is demonstrated in figures 1 (b) and (c). However, the refinement with co-existing *Cm* and *Pm* phases results in a more precise fit. Thus, the lowest goodness of fit ($\chi^2 = 4.06$) was obtained using the co-existing *Cm* + *Pm* phases. The coexistence of *Cm* and *Pm* phases in PMN-PT was earlier mentioned by *Singh et al.* [18] though others reported only the C*m* phase [20,21]. Earlier studies demonstrated that the phase stability in $(1 - x)PMN - xPT$ is dependent on *PT* composition ($x$). It was observed that single phase rhombohedral *R3m* structure was stable for $x < 0.27$, while tetragonal (*P4mm*) phase was more stable at $x > 0.39$. However, morphotropic phase boundary (MPB) region consists of coexisting *Cm* and *Pm* phases for 0.27 < x < 0.30, *Pm* and *P4mm* phases for 0.30 < x < 0.34, and *P4mm* and *Pm* phase for 0.34 < x < 0.39 [18]. Structural parameters of PMN-0.3PT estimated from each of the above refinements are listed in Table 1 and compared with previous reports. Earlier work by *Singh et al*. [22] reports the lattice parameters for PMN-0.29PT (*Cm*) and PMN-0.32PT (*Pm*) phases as *a* = 5.6968 Å, *b* = 5.6820 Å, *c* = 4.0132 Å and *a* = 4.0200 Å, *b* = 4.0046 Å, *c* = 4.0281 Å, respectively. The observed deviation of the lattice parameters with respect to previous literature could be attributed to structural distortion due to variation in processing conditions. The phase fraction obtained from the Rietveld refinement shows that the *Pm* is the dominant phase with a 73.81 % weight share in the present sample. Previous first-principles calculations also predicted monoclinic phase stability within the morphotropic phase boundary region and emphasized on the importance of the unstable polarization vector [23,24] to result in large polarization rotation. The rotation of the polarization vector for the $M_B$ ($Cm$) phase happens between the [101] and [111] directions, while the same occurs between the [001] and [101] directions for $M_c$ ($Pm$) phase, as shown schematically in Figure 2 (e). The rotation of the unstable polarization vector between the mentioned directions is mainly accountable for the large piezoelectric coefficients in the PMN-0.3PT piezoelectric material [23].

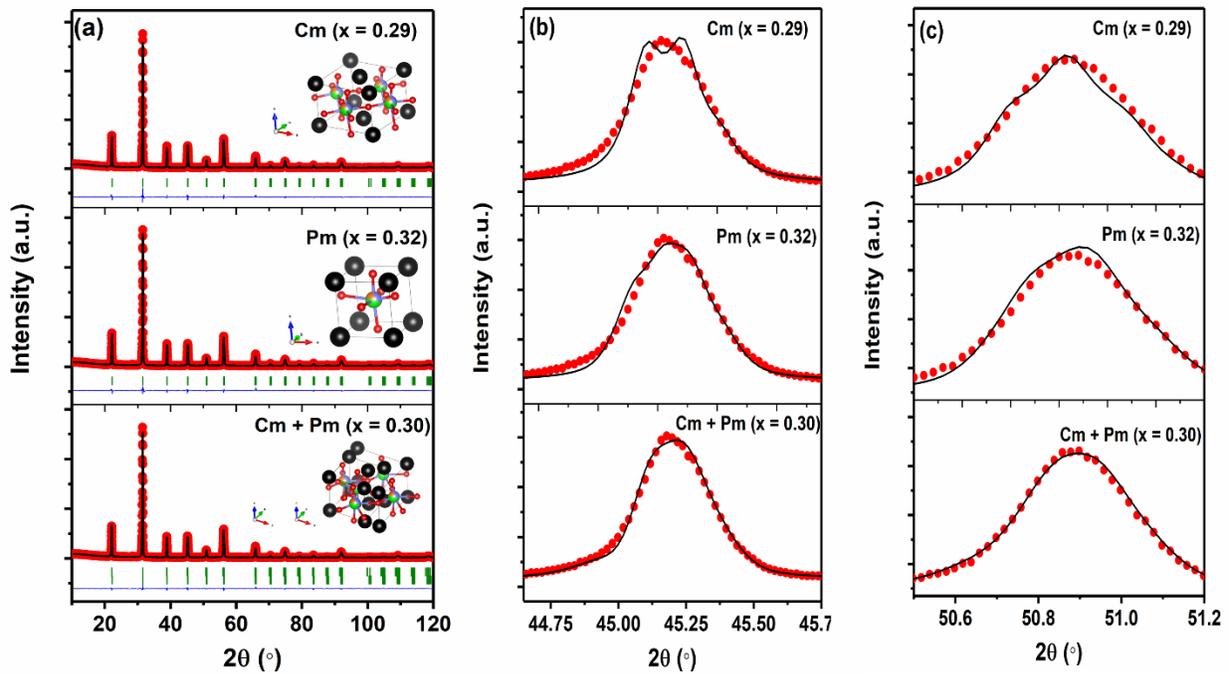

Figure 1. (a) Observed (dots), calculated (continuous line), Bragg position (vertical line), and difference (bottom line) profiles obtained by Rietveld refinement of the PMN-0.3PT XRD data using *Cm*, *Pm*, and *Cm + Pm* structural models (Inset: crystal structure with Pb in black, oxygen in red and Mg/Nb/Ti in orange/green/blue)) (b) Observed (dots), calculated (continuous line) profiles of (b) 200 (c) 210 reflections

Table 1. Structural parameters of monoclinic PMN-0.3PT estimated from the Rietveld refinement analysis of the powder x-ray diffraction (p-XRD) pattern.

| Parameters | *Cm* | *Pm* | *Cm + Pm* |
|---|---|---|---|
| *a* (Å) | 5.6811 ± 0.0003 | 4.0078 ± 0.0002 | 5.6642 ± 0.0010, 4.0086 ± 0.0002 |
| *b* (Å) | 5.6671 ± 0.0002 | 4.0000 ± 0.0002 | 5.6650 ± 0.0007, 4.0017 ± 0.0002 |
| *c* (Å) | 4.0009 ± 0.0002 | 4.0160 ± 0.0007 | 4.0302 ± 0.0003, 4.0158 ± 0.0001 |
| α (°) | 90 | 90 | 90, 90 |
| β (°) | 90.1244 | 90.1491 | 90.1471, 90.1258 |
| γ (°) | 90 | 90 | 90, 90 |
| $R_{wp}$ | 12.6 | 11.8 | 8.27 |
| $R_p$ | 12.8 | 11.9 | 8.75 |
| $R_{exp}$ | 4.12 | 4.12 | 4.10 |
| $\chi^2$ | 9.36 | 8.22 | 4.06 |
| Phase fraction (%) | | | *Cm* = 26.19, *Pm* = 73.81 |

Ionic coordinates within the unit cell allow one to estimate cation-oxygen polyhedral distortion. Octahedral distortion can be quantified by distortion index (D) [25], quadratic elongation ($\lambda$) and bond angle variance ($\sigma^2$) [26]. Distortion index represents the deviations of the Mg/Nb/Ti-O distances from their mean value. In an ideal (undistorted) octahedron, all the Mg/Nb/Ti-O distances are the same, *i.e.,* zero distortion index. Bond angle variance represents the deviation of O–Mg/Nb/Ti–O bond angle within the $BO_6$ (B = Mg/Nb/Ti) octahedron from 90° (in the case of an ideal octahedron). Hence, for an ideal octahedron, the bond angle variance becomes zero [27]. The crystal structural parameters of the Mg/Nb/Ti-O Octahedron estimated using the Rietveld data are listed in Table 2. The structure factors obtained from the Rietveld refinement were used to study the electron density distribution vis-à-vis bonding character of the cation-oxygen bonds PMN-0.3PT ceramic. Figure 2 (a) - (d) shows the 2D charge density mapping of the monoclinic phase with *Cm* and *Pm* space groups on the (101) and (111), and (001) and (101) planes respectively. In figures 2 (a) - (d), contour lines are disappeared between Pb–O and Mg/Nb/Ti-O ions, suggesting that the bonding nature between Pb and O ion is ionic. Similar nature of bonding was reported between Ba and O in $BaTiO_3$ perovskite [28].

Table 2. Structural parameters of the Mg/Nb/Ti-O Octahedron obtained from Rietveld refinement compared with the literature [22]

| Parameter | *Cm* [22] | *Pm* [22] | *Cm+Pm* (Present work) |
|---|---|---|---|
| Average bond length (Å) | 2.0124 | 2.0132 | 2.0326, 2.0111 |
| Polyhedral volume (Å$^3$) | 10.8253 | 10.8077 | 10.7768, 10.7364 |
| Distortion index | 0.01543 | 0.02405 | 0.11787, 0.03296 |
| Quadratic elongation | 1.0027 | 1.0055 | 1.0418, 1.0095 |
| Bond angle variance (deg.$^2$) | 8.8257 | 15.8629 | 92.6404, 24.2292 |

The surface morphology of the thermally etched PMN-0.3PT pellet is shown in figure 2 (f). A reasonably dense microstructure with few pores is observed. The sinter-density of the PMN-0.3PT pellet, measured by the Archimedes method, was found to be 7.38 ± 0.16 gm/cm$^3$, which is 90.55 % of the true density of the material. The mean grain size of the sample measured using Image J software [29] was found to be 1.2 ± 0.04 µm. The grain size distribution of the PMN-0.3PT pellet sample is shown in the insets of figure 2 (f). The chemical composition of PMN-0.3PT was determined by using Energy Dispersive Spectroscopy (EDS) technique. EDS analysis of the samples confirmed the absence of any foreign elements. (Please refer to Table S1 in the supplementary section) Although the variations in composition from the stoichiometric composition are under allowable limits [30], they may lead to complex defect chemistry and equilibria that could potentially impact charge transport vis-à-vis the material's electrical and electromechanical properties.

## 3.2 Impedance, dielectric, and electric modulus study

Figure 3 (a) and (b) plot variation of real ($Z'$) and imaginary ($Z''$) components of impedance with frequency over the temperature range of 350 to 425 °C. Figure 3 (a) reveals that the real impedance ($Z'$) decreases with increasing temperature in the initial frequency region (up to 10 kHz) and, after that, consolidates. This behavior could be attributed to the release of space charge with increased frequency and temperature [31]. Furthermore, it indicates that the charge transport rises with temperature and frequency. Figure 3 (b) shows that at a higher temperature, *viz.*, above 350 °C, the imaginary component of impedance demonstrates Debye-type peaks [32]. The peak in $Z''$ was found to shift towards higher frequencies with temperature, suggesting decreasing relaxation time ($\tau = 1/2\pi f_{max}$, where $f_{max}$ is the relaxation frequency) [32]. Further, it is noticed that with rising temperatures, there is a peak broadening, and at higher temperatures, the curves merge and appear to be flat. This type of behavior suggests relaxation caused by defects in the sample [33].

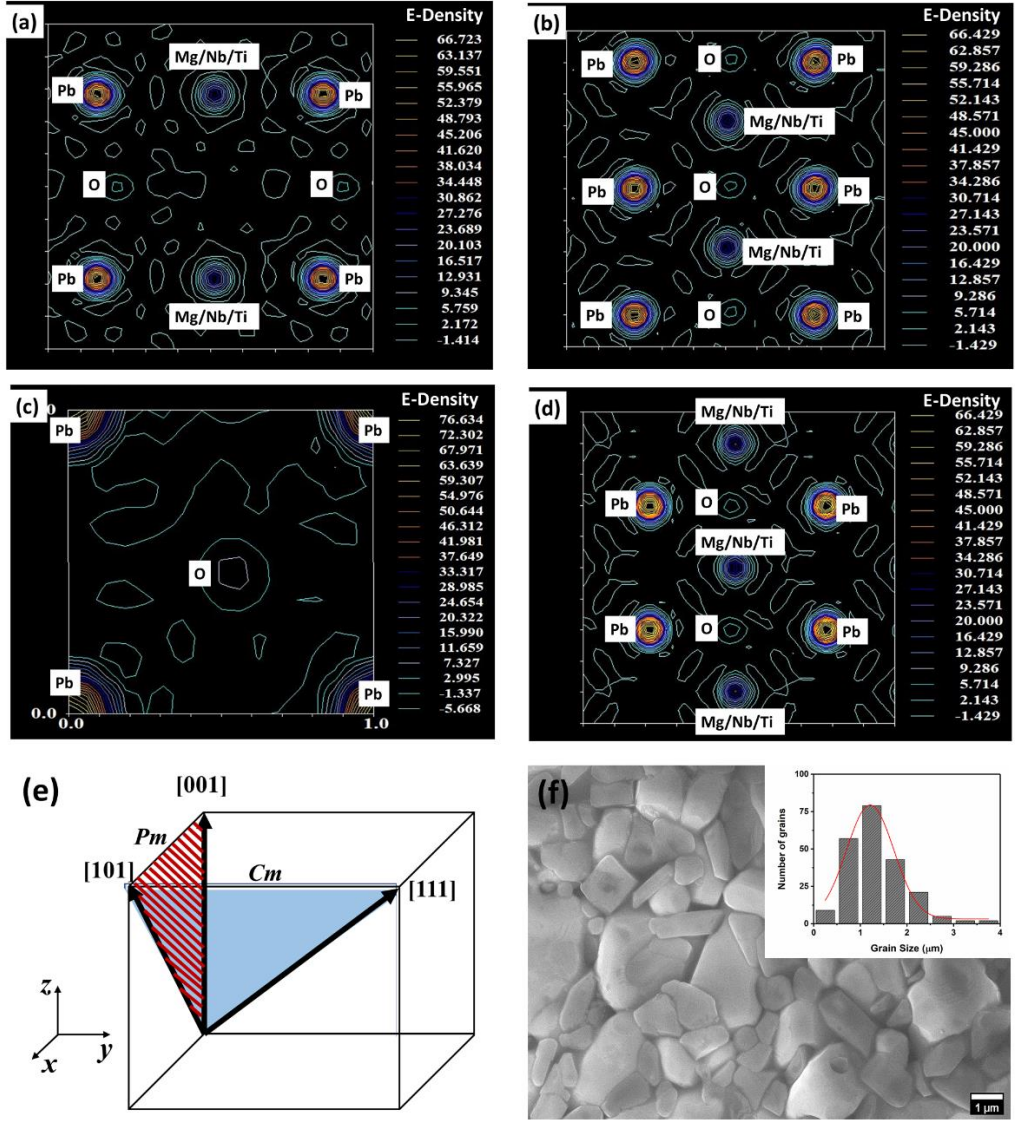

Figure 2. (a) Electron density distribution (2D) on (a) (101) plane for *Cm* phase (b) (111) plane for *Cm* phase (c) (001) plane for *Pm* phase (d) (101) plane for *Pm* phase (e) Polarization direction region for *Cm* and *Pm* (f) SEM micrograph of thermal-etched PMN-0.3PT pellet sample (Inset: grain size distribution)

Figure 3 (c) shows the Nyquist (Cole-Cole) plot of PMN-0.3PT ceramic pellet over a wide range of frequencies from 20 Hz to 1 MHz at 350 °C, 375 °C, 400 °C, and 425 °C. From the Nyquist plots, it is evident that the temperature variation has some noticeable response on the impedance spectrum. The appearance of double semicircular arcs emanating from the grain (bulk), grain boundary, and electrode in the conduction process [34]. The first semi-circle represents both grain and grain boundary contribution since the relaxation is not capacitance-dominated, as shown the Table 3. Further, the magnitude of capacitance (CPE) obtained by fitting ranges from $10^{-10}$ to $10^{-9}$ suggesting grain and grain boundary contribution [35]. However, the second semi-circle represents electrode contribution with a magnitude of capacitance (CPE) ranging from $10^{-8}$ to $10^{-7}$ [36]. The plot also reveals that grain and grain boundary relaxation are non-Debye type as the center of the semi-circle lies below the $Z'$ axis [37]. An increase in operating temperature causes a reduction in impedance. The low-frequency region represents electrode relaxation, whereas grain and grain boundary relaxation occur at a higher frequency. The semicircular arcs in the $Z'$ vs. $Z''$ plot was simulated from Z-view [38] by considering a parallel R-CPE circuit, where R and CPE are resistance and constant phase element [39]. The final obtained fitting parameters are shown in Table 3.

Table 3: Resistance and constant phase element as obtained from fitted data

| Temp (°C) | $R_g$ (Ω) | $CPE_g$ | $R_{gb}$ (Ω) | $CPE_{gb}$ | $R_e$ (Ω) | $CPE_e$ |
|---|---|---|---|---|---|---|
| 350 | 66540 | $1.6 \times 10^{-10}$ | $1.23 \times 10^5$ | $1.71 \times 10^{-10}$ | $7.62 \times 10^5$ | $1.83 \times 10^{-8}$ |
| 375 | 25540 | $2.05 \times 10^{-10}$ | 50540 | $2.1 \times 10^{-10}$ | $2.38 \times 10^5$ | $1.66 \times 10^{-8}$ |
| 400 | 7540 | $1.25 \times 10^{-10}$ | 15140 | $1.4 \times 10^{-10}$ | 62850 | $1.12 \times 10^{-8}$ |
| 425 | 2185 | $1.9 \times 10^{-9}$ | 5196 | $2.5 \times 10^{-9}$ | 15531 | $1.56 \times 10^{-7}$ |

Figures 3 (d) and (e) show the frequency dependence of real ($\kappa'$) and imaginary components of dielectric permittivity ($\kappa''$) at different temperatures (350 °C - 425 °C) calculated from the impedance data: [39]

$$\kappa' = \frac{t}{\omega \varepsilon_0 A} \times \frac{Z''}{(Z'^2 + Z''^2)} \tag{3}$$

$$\kappa'' = \frac{t}{\omega \varepsilon_0 A} \times \frac{Z'}{(Z'^2 + Z''^2)} \tag{4}$$

Where $t$ = sample thickness, $A$ = electrode area, $Z'$ = Real impedance, $Z''$ = imaginary impedance, $\kappa_0$ = permittivity of free space ($8.85 \times 10^{-12}$ F/m)

Both real ($\kappa'$) and imaginary ($\kappa''$) components of dielectric permittivity show a very large value at high temperature and low frequency; however, it appears to consolidate in the high-frequency region. High dielectric permittivity at low frequency is mainly attributed to interfacial polarization [40]. Both real and imaginary dielectric permittivity smoothly decreases with increasing frequency, attributed to the reduction in the polarization [41] due to thermal energy disrupting the dipole-dipole interaction and orientation. Figure 3 (f) shows the variation of dielectric permittivity with the temperature at the frequency range, 100 Hz to 1 MHz. The temperature-dependence dielectric permittivity plot showed a clear ferroelectric-paraelectric phase transition ($T_c$). It reveals relaxor-type ferroelectric behavior in PMN-0.3PT ceramic, as it doesn't exhibit a sharp phase transition at a $T_c$ [37].

Modulus study provides a complementary representation of the impedance characteristics of a material. The real and imaginary electric modulus are expressed in terms of permittivity as [39],

$$M' = \frac{\kappa'}{\left(\kappa'^2 + \kappa''^2\right)} \quad (5)$$

$$M'' = \frac{\kappa''}{\left(\kappa'^2 + \kappa''^2\right)} \quad (6)$$

Where $\kappa'$ and $\kappa''$ are the real and imaginary parts of dielectric permittivity, respectively.

The variation of real electric modulus ($M'$) with frequency at various temperatures from 350 °C to 425 °C is shown in figure 3 (g). The figure shows that the $M'$ is zero (approx.) at a lower frequency; however, $M'$ continually increases as the frequency increases and approaches the maximum at 1 MHz. These zero values of $M'$ at the lower frequency region are attributed to the absence of restoring force that can carry out the movement of charge carriers under the influence of the applied electric field. Within a low-frequency region, charges can move over long distances (long-range hopping) [42]. Furthermore, $M'$ increases with decreasing temperature, which can be attributed to a temperature-dependent relaxation mechanism [43]. The imaginary electric modulus ($M''$) versus frequency at different temperatures from 350 °C to 425 °C is shown in figure 3 (h). With increasing temperature, an increase in the peak height of the imaginary electrical modulus was observed due to the decrease in the capacitance with increasing temperature [42]. These asymmetric and broad peaks represent non-Debye type relaxation (different time constants), corroborating our conclusion of the impedance study. The figure shows the relaxation peak ($M''_{max}$) moves towards a higher frequency region with increased temperature. The low-frequency region below the relaxation peak represents the long-range hopping; however high-frequency region above the relaxation peak represents short-range hopping [44]. Such behavior indicates a temperature-dependent hopping process.

### 3.3 AC Conductivity

The ac conductivity ($\sigma_{ac}$) of the PMN-0.3PT sample was calculated using the eq. (7) given below [39]:

$$\sigma_{ac} = \frac{t}{A} \times \frac{Z'}{\left(Z'^2 + Z''^2\right)} \quad (7)$$

Where $t$ = sample thickness, $A$ = electrode area, $Z'$ = Real impedance, $Z''$ = imaginary impedance.

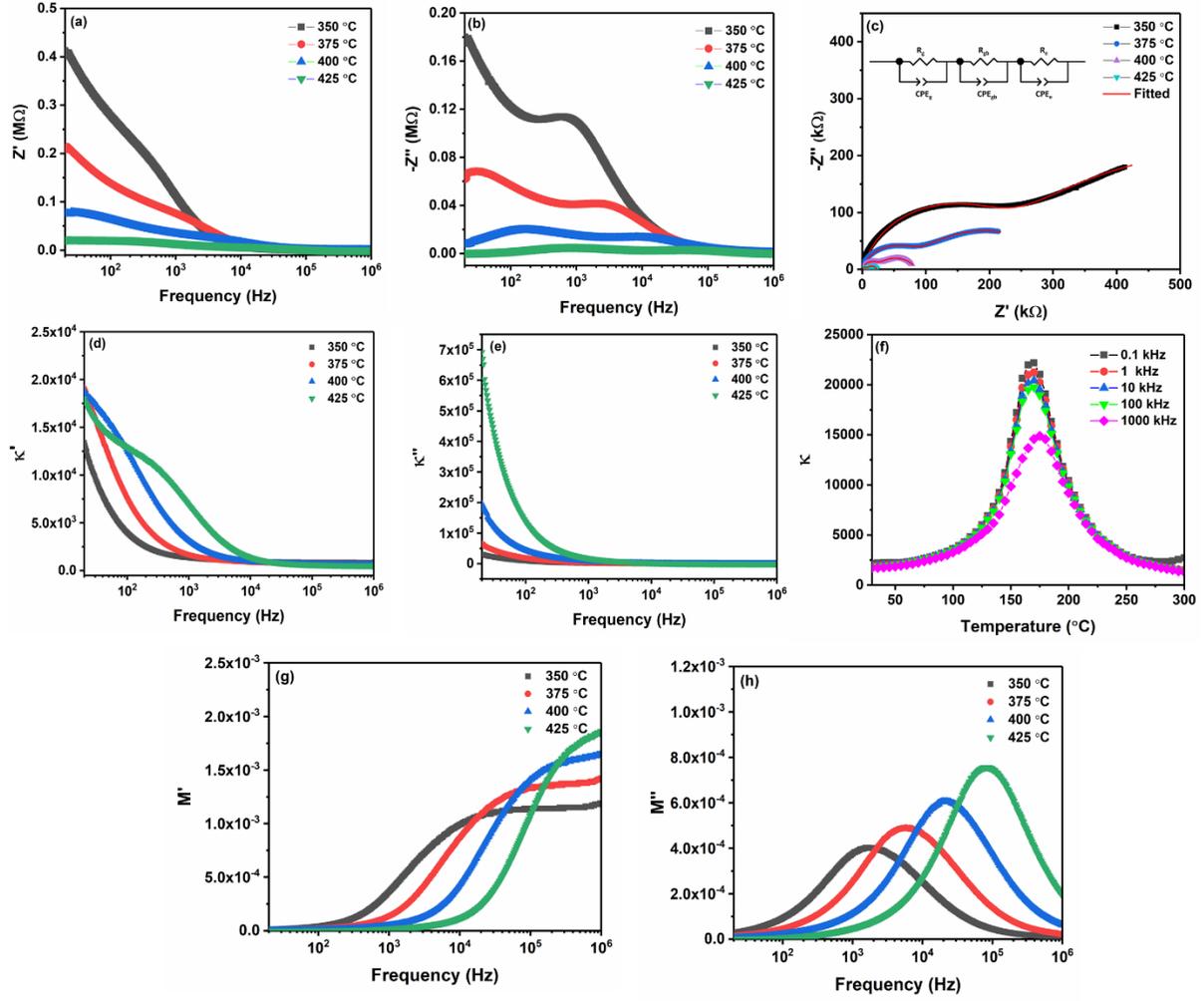

Figure 3. (a) Variation in the $Z'$ with frequency measured at different temperatures (b) Variation in the $Z''$ with frequency measured at different temperatures (c) Nyquist plot for PMN-0.3PT ceramic at different temperatures (d) Variation in the $\kappa'$ with frequency measured at different temperatures (e) Variation in the $\kappa''$ with frequency measured at different temperatures (f) Temperature dependence of dielectric constant at a different frequency (g) Variation in the $M'$ with frequency measured at different temperatures (h) Variation in the $M''$ with frequency measured at different temperature

Figure 4 (b) shows the ac conductivity ($\sigma_{ac}$) with $1000/T$ at 0.1 kHz, 1 kHz, and 10 kHz. The maximum ac conductivity (peak) was observed at around 170 °C at each frequency, as seen in the $\kappa$ vs. $T$ plot in figure 3 (f) [45]. This peak represents ferroelectric-paraelectric phase transition temperature. The ac conductivity at low temperatures shows both frequency and temperature-dependent; however, it almost merges at high temperatures [46]. There are different temperature ranges of varying conduction mechanisms: (a) low temperature (hopping charge conduction), (b) mid-temperature (oxygen vacancy conduction), and (c) high temperature (intrinsic ionic conduction) [47].

In PMN-0.3PT piezoceramic, there is a severe lead loss issue during pellet sintering at elevated temperatures. It is well-known that the ionization of oxygen will create oxygen vacancy [48]

$$O_O = V_O + \frac{1}{2}O_2 \tag{8}$$

Kröger–Vink notation of defects is used to describe the charge conduction, which is generated by the ionization of oxygen vacancies [49]

$$V_o = V_o^{\cdot} + e' \tag{9}$$

$$V_o^{\cdot} = V_o^{\cdot\cdot} + e' \tag{10}$$

where, $V_o^{\cdot}$ and $V_o^{\cdot\cdot}$ are singly-ionized and doubly-ionized oxygen vacancies, respectively. The defect scenario could be described using complementary defect reactions:

$$Pb_{Pb}^{\times} + O_o^{\times} = V_o^{\cdot\cdot} + V_{Pb}^{''} + PbO\ (\uparrow) \tag{11}$$

$$Pb_{Pb}^{\times} + O_o^{\times} = V_o^{\cdot} + V_{Pb}^{''} + PbO\ (\uparrow) + e' \tag{12}$$

$$Pb_{Pb}^{\times} + O_o^{\times} = V_o^{\times} + V_{Pb}^{''} + PbO\ (\uparrow) + 2e' \tag{13}$$

To balance the charge, oxygen vacancies formed in PMN-0.3PT ceramics, also known as extrinsic vacancies [48]. These high mobile oxygen vacancies can easily hop in the applied high electric field.

The variation of ac conductivity with the different temperature ranges indicates thermally activated conduction, which follows Arrhenius behavior, *i.e.*, $\sigma = \sigma_0 exp\ (-E_a/kT)$, where $k =$ Boltzmann's constant. Generally, oxygen vacancies are treated as the most movable charges and play an important role in conduction [50]. Figure 4 (a) suggests a frequency-independent dc conduction mechanism in the sample at high temperatures [51]. The activation energy for higher temperature regions (240 °C-350 °C) obtained at 0.1 kHz, 1kHz, and 10 kHz is 0.95 eV, 0.95 eV, and 1 eV, respectively, suggests frequency-independent activation energy [46]. The activation energies approx. 0.95 eV at higher temperature (≥ 240 °C) is because of the motion of doubly ionized oxygen vacancies rather than lead vacancies. The conduction due to Pb vacancies requires very high activation energy, close to 2 eV [46,52]. However, activation energy obtained at intermediate temperature (85 °C-145 °C) are 0.39 eV, 0.43 eV, and 0.43 eV at 0.1 kHz, 1 kHz, and 10 kHz frequency, respectively, suggesting that the mechanism of conduction is because of the motion of single ionized oxygen vacancies [53]. The activation energy for lower temperature regions (30 °C– 45 °C) obtained at 0.1 kHz, 1 kHz, and 10 kHz is 0.054 eV, 0.010 eV, and 0.004 eV respectively. This low activation energy and decreasing trend of activation energy with frequency suggest the conduction mechanism is due to the hopping of charge carriers. Therefore, low energy is needed for the charge carriers for the conduction [45].

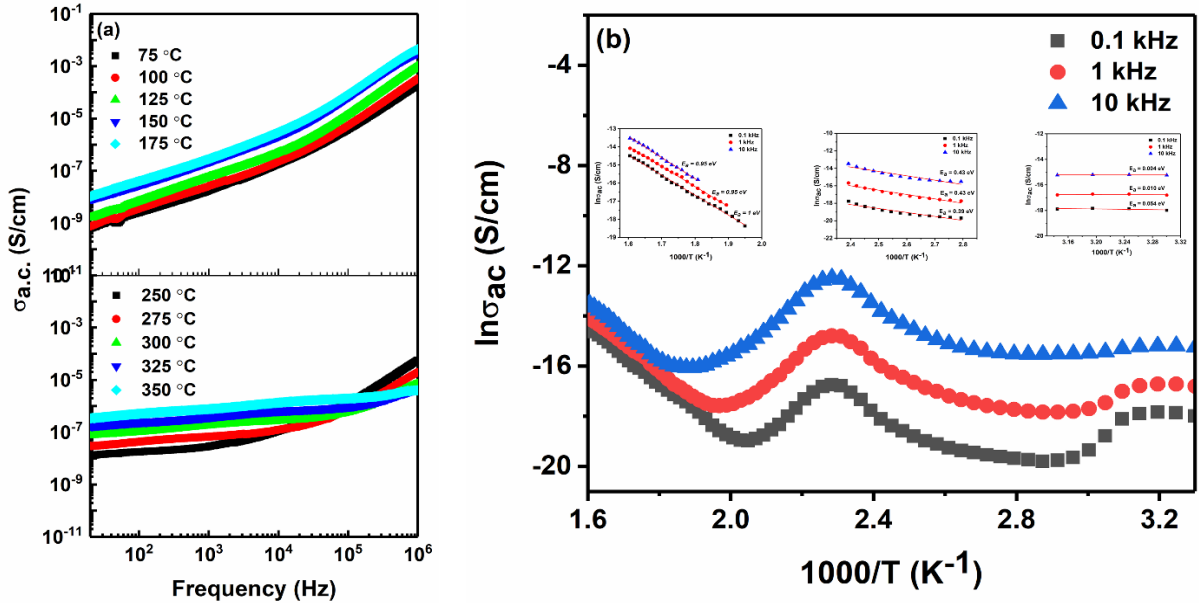

Figure 4. (a) ac conductivity plot vs. the frequency at different temperatures (b) Variation in the ac conductivity with 1000/T at different frequencies of PMN-0.3PT sample

Charged ionic and electronic defects in piezoceramics are detrimental since polarization leakage deteriorates the piezoelectric properties [54]. Thus, it is usual to minimize the concentration of mobile charge carriers by adding extra Pb to compensate for the loss [55] and/or doping with lanthanides [56]. Such a strategy could effectively reduce the concentration of oxygen vacancies and produce fewer mobile defects.

### 3.4 Electromechanical properties

The piezoelectric coefficient of the poled PMN-0.3PT samples was estimated using impedance measurement following the protocol outlined in European standard EN 50324-1:2002 and CEI/IEC 60483:1976 [57,58]. A cylindrical PMN-0.3PT sample ($t$ =10 mm, $d$ = 3 mm) was prepared following the above standard (inset of figure 5) and was poled at 10 kV for 30 min. at 80°C. The sample resonance ($f_r$) and anti-resonance frequency ($f_a$) were determined from the impedance vs. frequency plot. The resonance frequency of the piezoelectric materials is the frequency at which impedance is lowest, whereas anti-resonance frequency is the frequency at which impedance is highest. Electromechanical coupling factor ($k_{33}$), elastic compliance ($s_{33}^E$), dielectric permittivity ($\kappa_{33}^T$), and piezoelectric coefficient ($d_{33}$) were calculated using the following expressions and magnitudes are listed in Table 4 [59]:

$$s_{33}^D = \frac{1}{4 \times \rho \times f_a^2 \times t^2} \tag{14}$$

$$k_{33}^2 = \frac{\pi}{2} \times \frac{f_r}{f_a} \times \tan(\frac{\pi}{2} \times \frac{f_a - f_r}{f_a}) \tag{15}$$

$$s_{33}^E = \frac{s_{33}^D}{1-k_{33}^2} \tag{16}$$

$$\kappa_{33}^T = C^T \times \frac{t}{A}, \text{ where } A = \frac{\pi}{4} \times d^2 \tag{17}$$

$$d_{33} = k_{33} \, (\kappa_{33}^T \times s_{33}^E)^{\frac{1}{2}} \tag{18}$$

Where $\rho, t, A,$ and $d$ are the sample's density, thickness, electrode area, and diameter, respectively.

Resonance spectra of the longitudinal length mode of the PMN-0.3PT cylindrical sample are shown in figure 5. The piezoelectric coefficient estimated is $d_{33}$ ~ 200 ± 21 pC/N comparable to ~ 237 pC/N measured by a $d_{33}$ piezometer. The measured piezoelectric coefficient shows deviation from the available literature due to the low density of our sample [60–62] and loss of polarization current due to the presence of mobile charged defects. Similarly, elastic compliance and electromechanical coupling coefficient of the PMN-0.3PT were found to be less than 17.2 ×10⁻¹² m²/N and 0.61 in fine-grained PMN-0.3PT ceramic, 67.7 ×10⁻¹² m²/N and 0.92 in single-crystal PMN-0.3PT [63,64], respectively.

Table 4. Material coefficients determined using the resonance method for the longitudinal mode

| Material Coefficients | Value |
| --- | --- |
| $k_{33}$ | 0.54 ± 0.04 |
| $s_{33}^E$ [×10⁻¹² m²/N] | 13.16 ± 0.2 |
| $\kappa_{33}^T$ [× 10⁻⁸ F/m] | 1.06 ± 0.06 |
| $d_{33}$ [×10⁻¹² C/N] | 200 ± 21 |

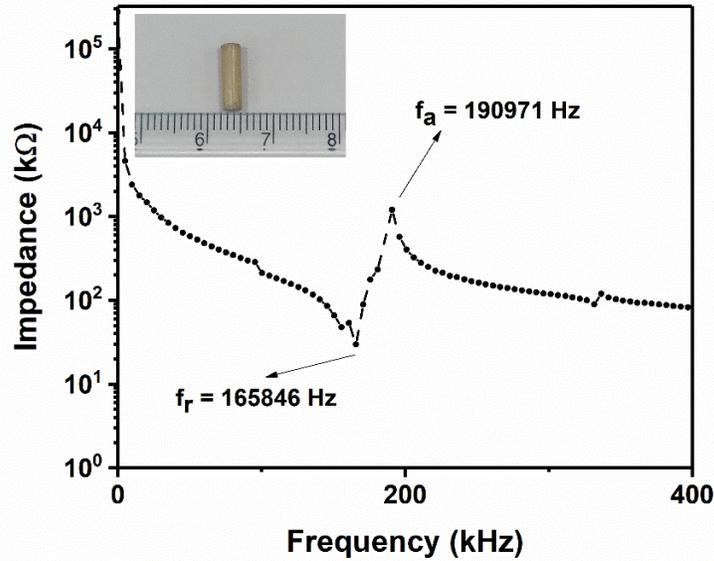

Figure 5. Resonance spectra of the longitudinal mode of the PMN-0.3PT sample

## 3.5 Energy harvesting

The energy harvesting performance of the PMN-0.3PT sample was evaluated by applying a cyclic, compressive load with an average amplitude of 3 N corresponding to a stress of 0.02 MPa, as shown in figure 6 (a). When subjected to a cyclic, compressive load, the short circuit current density and the open circuit electric field generated are shown in figures 6 (b) and (c). The maximum (mean) short circuit current density and maximum (mean) open circuit electric field generated from the PMN-0.3PT pellet is 95 (55) nA/cm$^2$ and 98 (82) V/cm, respectively. In output power density, 9.3 µW/cm$^3$ was generated from the present device, comparable to 11.05 µW/cm$^3$ in a vibration-based Mn-doped PZT energy harvester, but lesser than ~29 µW/cm$^3$ for PMN-0.25PT piezoelectric energy harvester due to off-resonance measurement in the present case [17]. Rakbamrung et al. [17] compared the performance of vibration energy harvester based on PMN-0.25PT and Mn-doped PZT and found 160 % higher output power in PMN-0.25PT piezoceramic. Chen et al. [65] developed polycrystalline PMN-PT-based cantilever energy harvesters and compared their results with similar type PZT harvesters. Their experimental results confirmed superior performance in PMN-PT-based harvesters. Roscow et al. [66] studied the piezoelectric performance of barium titanate pellet samples excited by a mechanical shaker. They reported an open-circuit voltage and short circuit current of 0.62 V and 360 nA, respectively, an estimated power density is much lower than the PMN-0.3PT.

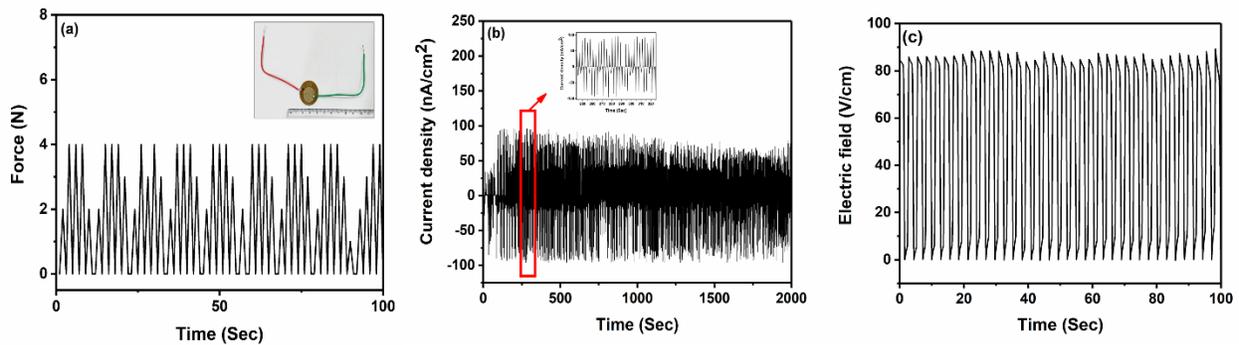

Figure 6. (a) Compressive cyclic load applies to the pellet sample (b) Current density response from PMN-0.3PT sample (b) Electric field response from PMN-0.3PT sample

## 3.6 Mechanical properties

Piezoceramics are notorious for their brittleness and thus poor flexibility. Therefore, harvesting vibrational energy in the compression mode using a constant amplitude, cyclic mechanical load demands assessing the material's mechanical stability. In this regard, we assess the fracture toughness and flexibility of the PMN-0.3PT in the present work.

Usually, piezoelectric ceramic tensile strength is very low compared to compressive strength due to internal defects (or pores). These defects act as stress concentrators. Furthermore, the fracture toughness of the piezoelectric ceramics is very low, restricting them in most applications [67] involving mechanical loading. Vickers hardness ($HV$) of the PMN-0.3PT ceramic, measured under a 98 mN, was found to be 6.37 ± 0.21 GPa, higher than 4.10 ± 0.5 GPa reported by Kathavate et al. [68]. The fracture toughness of PMN-0.3PT ceramic pellets was calculated from the crack-length method [69] after Vickers indentation testing. Many indentation models are developed based on the type of cracks, such as median crack and Palmqvist crack. Anstis [70] equation was used to calculate the fracture toughness of the PMN-0.3PT pellet as the cracks in the sample are of median type, as shown in figure 7. The equation for calculation of fracture toughness proposed by Anstis et al. is:

$$K_{IC} = 0.016 \left(\frac{E}{H}\right)^{0.5} \left(\frac{P}{c^{1.5}}\right) \tag{19}$$

The expression of fracture toughness proposed by Nihara et al. as per ASTM F 2094 is given by:

$$K_{IC} = \frac{10.3 \times E^{0.4} \times P^{0.6} \times a^{0.8}}{c^{1.5}} \tag{20}$$

Where $E$ is Young's modulus of PMN-0.3PT (76 GPa) obtained by using the resonance technique, $P$ is the indentation load (N), $H$ is the hardness (GPa), and $c$ is the crack length (m). A micrograph of typical Vickers indentation cracks on a PMN-0.3PT pellet at a load of 980 mN is shown in figure 7. It shows no such major cracks initiated from the edge of the indentation, only the cracks generated from the corners of the indentation can be observed. The fracture toughness of the PMN-0.3PT pellet samples was determined by eq. 19 and 20, found to be 0.58 ± 0.14 MPa.m$^{1/2}$ and 0.55 ± 0.13 MPa.m$^{1/2}$, respectively, which agree with the available literature [71]. The measured fracture toughness of the PMN-0.3PT ceramics is larger than most promising lead-free piezoceramic BZT-BCT i.e. 0.45 MPa.m$^{1/2}$ [72], BT (0.49 MPa.m$^{1/2}$) [73], however smaller than KNN-BLT-6BZ (0.70 MPa.m$^{1/2}$) [74], soft PZT (0.78 MPa.m$^{1/2}$) [75].

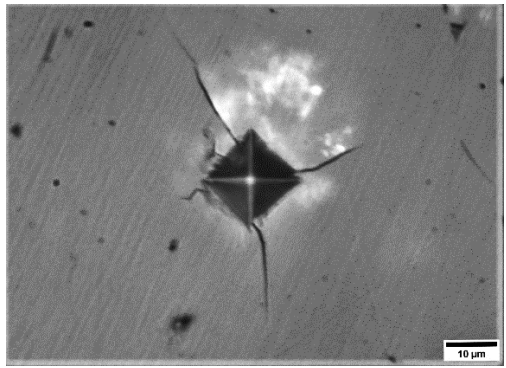

Figure 7. Cracks propagating in the PMN-0.3PT samples from a Vickers micro-indentation testing

The flexibility of piezoelectric materials is essential in assessing their usability in energy harvesting applications. It is always necessary to ensure that the piezoelectric material has high flexibility. Usually, flexible material should have a large elastic deformation, low elastic modulus, and, more importantly, can withstand cyclic compression loading for a long time [76,77]. These qualitative descriptions suggest that flexible materials possess a large yield strain ($\varepsilon_y$) [77]. Thus, yield strain is used to determine the figure-of-merit of the flexibility of materials: $f_{FOM} = \frac{\sigma_y}{E}$ [78] where $\sigma_y$, and $E$ are the yield stress and elastic modulus, respectively. The yield strength of the PMN-PT sample was taken as 34 MPa [79]. However, elastic moduli ($E_{33}$) of PMN-0.3PT ceramic was measured using resonance technique, $E \sim 7.6 \pm 0.11 \times 10^{10}$ N/m$^2$ comparable to $7.0 \times 10^{10}$ N/m$^2$ [80]. The flexibility figure-of-merit ($f_{FOM}$) of the PMN-0.3PT was calculated, found to be $4.5 \times 10^{-4}$, comparable to PZT ($5.7 \times 10^{-4}$) [81]. Despite having very high piezoelectric performance in the PMN-0.3PT piezoceramics, it is poor from actual service life. This has prompted the preparation of PGs based on piezoceramic-polymer composite, wherein the ceramic provides the attributes pertaining to piezoelectric coefficient while the polymer results require flexibility.

## 4. Conclusion

Pyrochlore-free PMN-0.3PT powder was synthesized using the solid-state reaction method. Rietveld refinement of p-XRD demonstrated the coexistence of two polar, monoclinic phases (Cm and Pm). The material demonstrated a $d_{33} \sim 200 \pm 21$ pC/N and a $T_c \sim 170$ °C. The reduced piezoelectric coefficient is attributed to the leakage of polarization current by the mobile oxygen vacancies, predicted by thorough impedance analyses. The estimated dielectric permittivity ($\kappa_{33}^T$), 1.06 ($\pm$ 0.06) $\times 10^{-8}$ F/m, and elastic modulus, 7.6 ($\pm$ 0.11) $\times 10^{10}$ N/m$^2$, are comparable to previous literature. The maximum short-circuit current density and the maximum electric field generated from the PMN-0.3PT piezoceramic are 95 nA/cm$^2$ and 98 V/cm, respectively, comparable to the available literature. However, the power density could possibly be improved by controlling the defect chemistry. The flexibility of the material estimated using yield strain, $4.5 \times 10^{-4}$, was found to be relatively small and not ideal for sustained cyclic loading. Texturing and/or forming polymer composite could be more suitable in this regard.

**Acknowledgments**

This work was funded by CSIR-HRDG, Govt. of India through project number, 22(0837)/20/EMR-II. Authors thank CIF, IIT Bhubaneswar for XRD and FESEM measurement.

**References**

[1]     Howells C A 2009 Piezoelectric energy harvesting *Energy Convers. Manag.* **50** 1847–50

[2]     Zeng W, Shu L, Li Q, Chen S, Wang F and Tao X M 2014 Fiber-based wearable electronics: A review of materials, fabrication, devices, and applications *Adv. Mater.* **26** 5310–36

[3]     Vullers R J M, van Schaijk R, Doms I, Van Hoof C and Mertens R 2009 Micropower energy harvesting *Solid. State. Electron.* **53** 684–93

[4]     Li H, Tian C and Deng Z D 2014 Energy harvesting from low frequency applications using piezoelectric materials *Appl. Phys. Rev.* **1**

[5]     Wang W, Yang T, Chen X and Yao X 2012 Vibration energy harvesting using a

piezoelectric circular diaphragm array *IEEE Trans. Ultrason. Ferroelectr. Freq. Control* **59** 2022–6

[6]  Kim H W, Priya S, Uchino K and Newnham R E 2005 Piezoelectric energy harvesting under high pre-stressed cyclic vibrations *J. Electroceramics* **15** 27–34

[7]  Xu T B, Siochi E J, Kang J H, Zuo L, Zhou W, Tang X and Jiang X 2013 Energy harvesting using a PZT ceramic multilayer stack *Smart Mater. Struct.* **22**

[8]  Toprak A and Tigli O 2014 Piezoelectric energy harvesting: State-of-the-art and challenges *Appl. Phys. Rev.* **1**

[9]  Covaci C and Gontean A 2020 Piezoelectric energy harvesting solutions: A review *Sensors (Switzerland)* **20** 1–37

[10]  Xu R and Kim S G 2012 FIGURES OF MERITS OF PIEZOELECTRIC MATERIALS IN ENERGY

[11]  Islam R A and Priya S 2006 Realization of high-energy density polycrystalline piezoelectric ceramics *Appl. Phys. Lett.* **88** 1–3

[12]  Mathers A, Moon K S and Yi J 2009 A vibration-based PMN-PT energy harvester *IEEE Sens. J.* **9** 731–9

[13]  Hwang G T, Park H, Lee J H, Oh S, Park K Il, Byun M, Park H, Ahn G, Jeong C K, No K, Kwon H, Lee S G, Joung B and Lee K J 2014 Self-powered cardiac pacemaker enabled by flexible single crystalline PMN-PT piezoelectric energy harvester *Adv. Mater.* **26** 4880–7

[14]  Tang G, Yang B, Liu J Q, Xu B, Zhu H Y and Yang C S 2014 Development of high performance piezoelectric d33 mode MEMS vibration energy harvester based on PMN-PT single crystal thick film *Sensors Actuators, A Phys.* **205** 150–5

[15]  Alomari A, Batra A, Aggarwal M and Bowen C R 2016 A multisource energy harvesting utilizing highly efficient ferroelectric PMN-PT single crystal *J. Mater. Sci. Mater. Electron.* **27** 10020–30

[16]  Chen Y, Zhang Y, Zhang L, Ding F and Schmidt O G 2017 Scalable single crystalline PMN-PT nanobelts sculpted from bulk for energy harvesting *Nano Energy* **31** 239–46

[17]  Rakbamrung P, Lallart M, Guyomar D, Muensit N, Thanachayanont C, Lucat C, Guiffard B, Petit L and Sukwisut P 2010 Performance comparison of PZT and PMN-PT piezoceramics for vibration energy harvesting using standard or nonlinear approach *Sensors Actuators, A Phys.* **163** 493–500

[18]  Singh A K and Pandey D 2003 Evidence for MB and MC phases in the morphotropic phase boundary region of (1-x)[Pb(Mg1/3Nb2/3)O3]-xPbTiO3: A Rietveld study *Phys. Rev. B - Condens. Matter Mater. Phys.* **67** 641021–6410212

[19]  Toby B H 2006  R factors in Rietveld analysis: How good is good enough? *Powder Diffr.* **21** 67–70

[20]  Hou D, Usher T M, Fulanovic L, Vrabelj M, Otonicar M, Ursic H, Malic B, Levin I and Jones J L 2018 Field-induced polarization rotation and phase transitions in 0.70Pb(M g1/3

N b2/3) O3-0.30PbTi O3 piezoceramics observed by in situ high-energy x-ray scattering *Phys. Rev. B* **97** 1–9

[21]   Singh C, Thakur V N and Kumar A 2021 Investigation on barometric and hydrostatic pressure sensing properties of Pb[(Mg1/3Nb2/3)0.7Ti0.3]O3 electro-ceramics *Ceram. Int.* **47** 6982–7

[22]   Singh A K, Pandey D and Zaharko O 2006 Powder neutron diffraction study of phase transitions in and a phase diagram of (1-x) [Pb (Mg1 3 Nb2 3) O3] -xPbTi O3 *Phys. Rev. B - Condens. Matter Mater. Phys.* **74** 1–18

[23]   Fu H and Cohen R E 2000 Polarization rotation mechanism for ultrahigh electromechanical response in single-crystal piezoelectrics *Nature* **403** 281–3

[24]   Vanderbilt D and Cohen M H 2001 Monoclinic and triclinic phases in higher-order Devonshire theory *Phys. Rev. B - Condens. Matter Mater. Phys.* **63** 1–9

[25]   Baur W H 1974 The geometry of polyhedral distortions. Predictive relationships for the phosphate group *Acta Crystallogr. Sect. B Struct. Crystallogr. Cryst. Chem.* **30** 1195–215

[26]   Robinson K, Gibbs G V. and Ribbe P H 1971 Quadratic Elongation: A Quantitative Measure of Distortion in Coordination Polyhedra *Science* **172** 567–70

[27]   Ben Hamed H, Hoffmann M, Adeagbo W A, Ernst A, Hergert W, Hynninen T, Kokko K and Paturi P 2019 First-principles investigations of the magnetic phase diagram of Gd1-xCaxMnO3 *Phys. Rev. B* **99** 1–11

[28]   Sasikumar S, Thirumalaisamy T K, Saravanakumar S, Asath Bahadur S, Sivaganesh D and Shameem Banu I B 2020 Effect of neodymium doping in BaTiO3 ceramics on structural and ferroelectric properties *J. Mater. Sci. Mater. Electron.* **31** 1535–46

[29]   Nath A K and Medhi N 2012 Density variation and piezoelectric properties of Ba(Ti 1-xSnx)O3 ceramics prepared from nanocrystalline powders *Bull. Mater. Sci.* **35** 847–52

[30]   Bochenek D, Skulski R and Niemiec P 2018 Electrophysical properties of the PMN-PT-PS solid solution *Materials (Basel).* **11** 1–14

[31]   Choudhary R N P, Pradhan D K, Tirado C M, Bonilla G E and Katiyar R S 2007 Structural, dielectric and impedance properties of Ca(Fe2/3W1/3)O3 nanoceramics *Phys. B Condens. Matter* **393** 24–31

[32]   Srivastava A, Garg A and Morrison F D 2009 Impedance spectroscopy studies on polycrystalline BiFeO3 thin films on Pt/Si substrates *J. Appl. Phys.* **105**

[33]   Yan X, Chen X, Li X, Liu G, Zhang H and Zhou H 2018 Good electrical performances and impedance analysis of (1 − x)KNN–xBMM lead-free ceramics *J. Mater. Sci. Mater. Electron.* **29** 4538–46

[34]   Sethi G, Bontempo B, Furman E, Horn M W, Lanagan M T, Bharadwaja S S N and Li J 2011 Impedance analysis of amorphous and polycrystalline tantalum oxide sputtered films *J. Mater. Res.* **26** 745–53

[35]   Irvine B J T S, Sinclair D C and West A R 1990 Electroceramics Characterisation by

impedance sepctroscopy.pdf *Adv. Mater.* **2** 132–8

[36]  Hayat K, Ali S, Rahman A U, Khan S, Shah S K and Iqbal Y 2018 Effect of B-site dopants on the electrical properties of BaMn1-xAxO3 ceramics via low temperature impedance spectroscopy *Mater. Res. Express* **5**

[37]  Subohi O, Bowen C R, Malik M M and Kurchania R 2016 Dielectric spectroscopy and ferroelectric properties of magnesium modified bismuth titanate ceramics *J. Alloys Compd.* **688** 27–36

[38]  Johnson D 2000 Software Zview-v 3.1 *Scribner Assoc. Inc.*

[39]  Joshi J H, Kanchan D K, Joshi M J, Jethva H O and Parikh K D 2017 Dielectric relaxation, complex impedance and modulus spectroscopic studies of mix phase rod like cobalt sulfide nanoparticles *Mater. Res. Bull.* **93** 63–73

[40]  Rayssi C, El Kossi S, Dhahri J and Khirouni K 2018 Frequency and temperature-dependence of dielectric permittivity and electric modulus studies of the solid solution Ca0.85Er0.1Ti1-: XCo4 x /3O3 (0 ≤ x ≤ 0.1) *RSC Adv.* **8** 17139–50

[41]  Barick B K, Mishra K K, Arora A K, Choudhary R N P and Pradhan D K 2011 Impedance and Raman spectroscopic studies of (Na0.5Bi 0.5)TiO3 *J. Phys. D. Appl. Phys.* **44**

[42]  Rehman F, Li J B, Ahmad I, Jin H B, Ahmad P, Saeed Y, Shafiq M, Ali H, Khan M I and Ali T 2019 Dielectric relaxation and electrical properties of Bi 2.5 Nd 0.5 Nb 1.5 Fe 0.5 O 9 ceramics *Mater. Chem. Phys.* **226** 100–5

[43]  Ben Bechir M, Karoui K, Tabellout M, Guidara K and Ben Rhaiem A 2018 Dielectric relaxation, modulus behavior and thermodynamic properties in [N(CH3)3H]2ZnCl4 *Phase Transitions* **91** 901–17

[44]  Brahma S, Choudhary R N P and Thakur A K 2005 AC impedance analysis of LaLiMo 2O 8 electroceramics *Phys. B Condens. Matter* **355** 188–201

[45]  Behera B, Arajo E B, Reis R N and Guerra J D S 2009 AC conductivity and impedance properties of 0.65Pb(Mg 1/3 Nb 2/3 )O 3 -0.35 PbTiO 3 ceramics *Adv. Condens. Matter Phys.* **2009**

[46]  Pandit P and Bangotra P 2021 AC conductivity and dielectric analysis of Pb1-xLax[(Mg1+x/3Nb2-x/3)0.65Ti0.35(1-x)/4]O3, (x = 0, 0.02, 0.05) relaxor ferroelecric ceramics *Adv. Mater. Proc.* **1** 131–9

[47]  Raymond O, Font R, Suárez-Almodovar N, Portelles J and Siqueiros J M 2005 Frequency-temperature response of ferroelectromagnetic PbÑFe1/2Nb1/2ÖO3ceramics obtained by different precursors. Part I. Structuraland thermo-electrical characterization *J. Appl. Phys.* **97**

[48]  Zhang T F, Tang X G, Liu Q X, Lu S G, Jiang Y P, Huang X X and Zhou Q F 2014 Oxygen-vacancy-related relaxation and conduction behavior in (Pb1-xBax)(Zr0.95Ti0.05)O3 ceramics *AIP Adv.* **4** 0–11

[49]  Peliz-Barranco A, Guerra J D S, López-Noda R and Arajo E B 2008 Ionized oxygen

vacancy-related electrical conductivity in (Pb 1-xLax)(Zr0.90Ti0.10) 1-x/4O3 ceramics *J. Phys. D. Appl. Phys.* **41**

[50]   Kim J S and Kim I W 2005 Role of the doped Nb ion in ferroelectric bismuth titanate and bismuth lanthanum titanate ceramics: Different dielectric behaviors *J. Korean Phys. Soc.* **46** 143–6

[51]   Skulski R, Wawrzała P, KORZEKWA J and SZYMONIK M 2009 The electrical conductivity of PMN-PT ceramics *Arch. Metall. Mater.* **54** 935–41

[52]   Guiffard B, Boucher E, Eyraud L, Lebrun L and Guyomar D 2005 Influence of donor co-doping by niobium or fluorine on the conductivity of Mn doped and Mg doped PZT ceramics *J. Eur. Ceram. Soc.* **25** 2487–90

[53]   Peláiz-Barranco A and Guerra J D S 2010 Dielectric relaxation related to single-ionized oxygen vacancies in (Pb1-xLax)(Zr0.90Ti0.10) 1-x/4O3 ceramics *Mater. Res. Bull.* **45** 1311–3

[54]   Wang J, Shen Y G, Song F, Ke F J, Bai Y L and Lu C S 2016 Effects of oxygen vacancies on polarization stability of barium titanate *Sci. China Physics, Mech. Astron.* **59** 1–4

[55]   Kobor D, Hajjaji A, Garcia J E, Perez R, Albareda A, Lebrun L and Guyomar D 2010 Dielectric and Mechanical Nonlinear Behavior of Mn Doped PMN-35PT Ceramics *J. Mod. Phys.* **01** 211–6

[56]   Vendrell X, García J E, Cerdeiras E, Ochoa D A, Rubio-Marcos F, Fernández J F and Mestres L 2016 Effect of lanthanide doping on structural, microstructural and functional properties of K0.5Na0.5NbO3 lead-free piezoceramics *Ceram. Int.* **42** 17530–8

[57]   Anon 2002 European standard, EN 50324-1:2002 - Piezoelectric properties of ceramic materials and components - Part 1: Terms and definitions *CENELEC Eur. Comm. Electrotech. Stand.*

[58]   Anon 1997 International Standard, CEI/IEC 60483: 1976 : Guide to Dynamic Measurements of Piezoelectric Ceramics with High Electromechanical Coupling *IEE Geneva, Switz.* **p. 40**

[59]   Anon 2002 European standard, EN 50324-2:2002, "Piezoelectric properties of ceramic materials and components - Part 2: Methods of measurement – Low power" *CENELEC Eur. Comm. Electrotech. Stand.*

[60]   Zuo R, Granzow T, Lupascu D C and Rödel J 2007 PMN-PT ceramics prepared by spark plasma sintering *J. Am. Ceram. Soc.* **90** 1101–6

[61]   Pham-Thi M, Augier C, Dammak H and Gaucher P 2006 Fine grains ceramics of PIN-PT, PIN-PMN-PT and PMN-PT systems: Drift of the dielectric constant under high electric field *Ultrasonics* **44** 6–10

[62]   Qian K, Fang B, Du Q, Ding J, Zhao X and Luo H 2013 Phase development and electrical properties of Pb(Mg1/3Nb 2/3)O3-PbTiO3 ceramics prepared by partial oxalate route *Phys. Status Solidi Appl. Mater. Sci.* **210** 1149–56

[63]   Wang H, Jiang B, Shrout T R and Cao W 2004 Electromechanical Properties of Fine-Grain , 0 . 7 Pb ( Mg 1 / 3 Nb 2 / 3 ) O 3 -0 . 3PbTiO 3 Ceramics 1–4

[64]   Zhang R, Jiang W, Jiang B and Cao W 2002 Elastic, Dielectric and Piezoelctric Coefficients of Domain Engineered 0.70Pb(Mg1/3Nb2/3)O3-0.30PbTiO3 Single Crystal *AIP Conf. Proc.* **626**

[65]   Chen C T, Lin S C, Trstenjak U, Spreitzer M and Wu W J 2021 Comparison of metal-based pzt and pmn–pt energy harvesters fabricated by aerosol deposition method *Sensors* **21**

[66]   Roscow J I, Zhang Y, Kraśny M J, Lewis R W C, Taylor J and Bowen C R 2018 Freeze cast porous barium titanate for enhanced piezoelectric energy harvesting *J. Phys. D. Appl. Phys.* **51**

[67]   Lan C F, Nie H C, Chen X F, Wang J X, Wang G S, Dong X L, Liu Y S and He H L 2013 Effects of poling state and pores on fracture toughness of Pb(Zr 0·95Ti0·05)O3 ferroelectric ceramics *Adv. Appl. Ceram.* **112** 306–10

[68]   Kathavate V S, Praveen Kumar B, Singh I and Eswar Prasad K 2020 Effect of sub and above-curie temperature annealing on the nanomechanical properties of PMN-PT piezoceramics *Ceram. Int.* **46** 12876–83

[69]   Evans A G and Charles E A 1976 Fracture Toughness Determination by indentation *J. Am. Ceram. Soc.* **59** 371–2

[70]   Anstis G R, Chantikul P, Lawn B R and Marshall D B 1981 A Critical Evaluation of Indentation Techniques for Measuring Fracture Toughness: I, Direct Crack Measurements *J. Am. Ceram. Soc.* **64** 533–8

[71]   Wang H and Singh R N 1995 Electric field effects on the crack propagation in an electrostrictive pmn-pt ceramic *Ferroelectrics* **168** 281–91

[72]   Vögler M, Acosta M, Brandt D R J, Molina-Luna L and Webber K G 2015 Temperature-dependent R-curve behavior of the lead-free ferroelectric 0.615Ba(Zr0.2Ti0.8)O3-0.385(Ba0.7Ca0.3)TiO3 ceramic *Eng. Fract. Mech.* **144** 68–77

[73]   Schneider G A, Meschke F, Raddatz O and Kolleck A 2000 R-Curve Behavior and Crack-Closure Stresses in Barium Titanate and (Mg,Y)-PSZ Ceramics *J. Am. Ceram. Soc.* **83** 353–61

[74]   Li Y, Liu Y, Öchsner P E, Isaia D, Zhang Y, Wang K, Webber K G, Li J F and Rödel J 2019 Temperature dependent fracture toughness of KNN-based lead-free piezoelectric ceramics *Acta Mater.* **174** 369–78

[75]   Vögler M, Fett T and Rödel J 2018 Crack-tip toughness of lead-free (1−x)(Na1/2Bi1/2)TiO3–xBaTiO3 piezoceramics *J. Am. Ceram. Soc.* **101** 5304–8

[76]   Wong W S, Chabinyc M L, Ng T-N and Salleo A 2009 *Materials and Novel Patterning Methods for Flexible Electronics*

[77]   Chen H, Wei T, Zhao K, Qiu P, Chen L, He J and Shi X 2021 Room-temperature plastic inorganic semiconductors for flexible and deformable electronics *InfoMat* **3** 22–35

[78]   Peng J and Jeffrey Snyder G 2019 A figure of merit for flexibility *Science (80-. ).* **366** 690–1

[79]   Ewart L M, McLaughlin E A, Robinson H C, Amin A and Stace J J 2007 Mechanical and electromechanical properties of PMN-PT single crystals for naval sonar transducers *IEEE Int. Symp. Appl. Ferroelectr.* 553–6

[80]   Shih W Y, Luo H, Li H, Martorano C and Shih W H 2006 Sheet geometry enhanced giant piezoelectric coefficients *Appl. Phys. Lett.* **89**

[81]   Fett T, Munz D and Thun G 1999 Tensile and bending strength of piezoelectric ceramics *J. Mater. Sci. Lett.* **18** 1899–902